# Toward a Research Software Security Maturity Model

Randy Heiland, Betsy Thomas, Von Welch, Craig Jackson
Center for Trustworthy Scientific Cyberinfrastructure, Indiana University
{heiland, fnubetsy, vwelch, scjackso}@indiana.edu

## Introduction

In its *Vision and Strategy for Software for Science, Engineering, and Education* [NSF-CIF21] the NSF states that it will invest in activities that:

> "Recognize that software strategies must include the secure and reliable deployment and operation of services, for example by campuses or national facilities or industry, where identity, authentication, authorization and assurance are crucial operational capabilities."

and

> "Result in high-quality, usable, secure, vulnerability-free, sustainable, robust, well-tested, and maintainable/evolvable software; and which promotes the sustainability of solid and useful on-going investments."

Such statements evidence that security should indeed be a first-class consideration of the software ecosystem. In this position paper, we share some thoughts related to research software security. Our thoughts are based on the observation that security is not a binary, all-or-nothing attribute, but a range of practices and requirements depending on how the software is expected to be deployed and used. We propose that the community leverage the concept of a maturity model, and work to agree on a research software security maturity model. This model would categorize different sets of security needs of the deployment community, and provide software developers a roadmap for advancing the security maturity of their software. The intent of this paper is not to express such a comprehensive maturity model, but instead to start a conversation and set some initial requirements.

## The Argument for a Maturity Model Approach

Software, depending on its expected deployment conditions, will have different requirements for security. For example, prototype software written by a graduate student, never intended to be run in other environments, may not address security at all. Software intended to be run as a service on production infrastructure will be expected to meet some level of security. Other sets of requirements exist between and outside of these points. It is reasonable to expect software intended for varied purposes to meet different standards of security.

In research computing, we need to support a range of requirements with regards for software security . However, development and deployment are often carried out by different people in different organizations and projects. This highlights the need for a common terminology and framework shared between software deployers and developers for what they require and provide respectively.

Brought together, a research software security maturity model would lay out levels of security requirements and the software engineering and maintenance practices that meet those requirements.

## Existing Software Security Maturity Models

There exist a number of examples in the commercial and open source space:

- BSIMM [BSIMM] is a maturity model which defines the Software Security Framework, which outlines twelve security practices (e.g., Strategy and Metrics, Attack Models, Architectural Analysis, Penetration Testing). The maturity model consists of a defined set of levels of activities within each area of practice.

- vBSIMM [vBSIMM] is a variant of BSIMM focused on organizations who receive software from other parties. It is a subset of five security practices out of twelve from the full BSIMM software security framework.

- OPENSAMM [OpenSAMM] is an open framework to aid organizations in evaluating their existing software security practices. OPENSAMM focuses on four core activities for software development present in any organization, namely, Governance, Construction, Verification, Deployment. For each of these functions, three security activities have been defined by outlining success metrics, personnel overhead, and associated costs. Each of these three activities are defined in increasing order of maturity.

## Existing Research Software Requirements

Below are notable examples of research software requirements:

- The NSF XSEDE project (www.xsede.org) has a SDI CI Security Considerations document [XSEDE-SDI] which consists of a checklist of security considerations such as Risks and Attack Surface, Authentication and Access Control, Software Security & Change Management, Network Security, Logging, Privacy, and Changes to Security Infrastructure.

- The Software Sustainability Institute Software Evaluation Guide [SSI] is a software evaluation checklist that contains items for code review, usability of software and sustainability (but not security however). This is the first step in a Sustainability Evaluations Guide [SSI2].

We note both examples lack any discussion of software security.

## Goals of A Research Software Security Maturity Model

A research software security maturity model should address the following goals:

1. It should provide software developers and maintainers a well-defined taxonomy for different sets of security requirements and the software development and maintenance processes that meet those requirements. This taxonomy would support software developers and maintainers in selecting clear targets for deployment, and support software deployers and users in selecting software that meets their needs.

2. It should provide software developers with a roadmap for increasing the security of their software. By providing clear levels of maturity, software developers would be guided in the appropriate next steps to advance their processes and, hence, the security of their software.

3. Given finite resources of research projects, the model should be relatively simple and easy to understand by software developers, such that the benefits of the prior two goals are not eliminated by the effort to understand the model.

4. By providing guidance and a roadmap to the software development community, the model should slowly raise the overall level of software security by making the process more understandable and tractable.

5. The model should be agnostic to the software development methodologies (e.g., Agile [agile]) in use by different projects.

## Recommendations for Next Steps

Determining a set of requirements for security as needed by deploying projects and categorizing those would be a necessary step towards a useful maturity model. This could be done incrementally, starting with a single set of requirements from a single project and then letting other sets be defined relative to the first. The expectation being these sets would naturally cluster over time into maturity levels.

Of the software maturity models described previously, OpenSAMM, with its open source origins, would seem to be a good starting point for the research community. We especially like OpenSAMM's position that software requirements be specific, measurable, and reasonable. In the long term, software developers will benefit from specific guidance and tools to help with obtaining different maturity levels; sharing a model with the open source community would help in leveraging their efforts in this regard.

We note that a security maturity model could be part of a general software maturity model, covering issues such as documentation, support, testing, and portability (e.g., [SSI, SSI2]). However, even a preliminary discussion of a more general model is outside the scope of this document.

## Alternatives to and Limitations of Maturity Models

There exist software security lifecycles (e.g., [MS-SDL]) that provide a fixed set of software security practices, and hence address a fixed set of requirements. However, given the range of developmental conditions that research software is expected to fall into, we suggest that a maturity model expressing of a range of security maturity levels is the appropriate approach for research software.

In terms of limitations, we believe security maturity models primarily address metrics focused on the development process rather than real world results (e.g., software proving itself more secure in deployment). However, we believe this is reasonable given the difficulty in both defining result-oriented metrics and objectively measuring them in a reasonable amount of time.

## Acknowledgment

This paper is a product of the Center for Trustworthy Scientific Cyberinfrastructure (CTSC) / trustedci.org, supported by the National Science Foundation under Grant Number OCI-1234408. Any opinions, findings, and conclusions or recommendations expressed in this material are those of the authors and do not necessarily reflect the views of the National Science Foundation.